\documentclass[11pt,a4paper]{article}
\usepackage{a4wide}
\usepackage[dvips]{graphics}

\usepackage{epsf}

\newcommand{\beq}{\begin{equation}}
\newcommand{\eeq}{\end{equation}}
\newcommand{\bea}{\begin{eqnarray}}
\newcommand{\eea}{\end{eqnarray}}
\newcommand{\ch}{{\mathcal{H}}}
\newcommand{\cR}{{\mathcal{R}}}
\newcommand{\base}{{\mathbf{\hat{e}}}}
\newcommand{\fd}{{\mathbf{d}}}

\begin{document}

\title{On Second Order Superhorizon Perturbations in Multi-Field Inflationary Models}
\author{Gerasimos Rigopoulos\\
{\it Department of Applied Mathematics and Theoretical Physics,}\\
{\it Centre for Mathematical Sciences,}\\
{\it University of Cambridge,}\\
{\it Wilberforce Road, Cambridge, CB3 0WA, UK.}}

\maketitle

\begin{abstract}

We present a method for the study of second
order superhorizon perturbations in multi field inflationary
models with non trivial kinetic terms. We utilise a change of coordinates in field space to separate
isocurvature and adiabatic perturbations generalizing previous results. We also construct second order gauge invariant
variables related to them. It is found that with an arbitrary metric in field space the isocurvature perturbation sources the gravitational
potential on long wavelengths even for ``straight''
trajectories. The potential decouples from the isocurvature
perturbations if the background fields' trajectory is a
geodesic in field space. Taking nonlinear effects into
account shows that, in general, the two types of perturbations couple to each other. This is
an outline of a possible procedure to study nonlinear
and non-Gaussian effects during multifield inflation.

\end{abstract}

\section {Introduction}
It has been customary to say that an adiabatic, gaussian, almost scale
invariant perturbation is a generic prediction of inflation. In the past few
years though it has been realised that if more than one
degrees of freedom are allowed to be relevant during inflation - i.e
more dynamic scalar fields - then isocurvature perturbations can arise,
possibly correlated with the adiabatic ones, leading to a far richer
phenomenology. For example simple statements about single field inflation such as
the conservation of the superhorizon curvature perturbation $\cR$ do
not hold in multifield models \cite{wands1}. Therefore the need for more accurate
modeling of the inflationary era has arisen and is actively pursued at
the moment. 

So far almost all studies of inflationary perturbations have been performed using only linear perturbation
theory. The smallness of the fluctuations in the temperature of the
CMB certainly justify this approach. But given the accuracy of the
forthcoming data it would be worth trying to go beyond this
approximation and see if we can extract more information about our
models by studying nonlinear effects during inflation. Nonlinearities
are always there since gravity is a nonlinear theory. They would
induce non gaussianities in the fluctuations of the cosmic
microwave background which could be potentially observable by the
PLANCK sattelite due for launch in 2007. The level of nongaussianity
in standard single field models of inflation has been estimated in the
past (e.g \cite{nogauss, sb, mald, mat} - ref. \cite{sakel} discusses
deviations from gaussianity within linear theory but from a non vacuum
initial state). It turns
out that such a signal will not be detectable even by PLANCK
\cite{kom}. 

Multiple scalar fields
seem to have a better chance of producing an observable nongaussian
signal \cite{yi,uzan} the detection of which could provide evidence that more
than one degrees of freedom were relevant during inflation. Current
limits from WMAP on the deviation of the CMB from gaussianity come in the
form of an allowed
range for a widely used non-linearity parameter $f_{NL}$ related to a
$\chi^2$ type of non gaussianity. The authors
of \cite{wmap} find $-58<f_{NL}<134$ (95\% C.L.). It would be interesting to see
if multiple fields can generate observable non gaussianity within this
limit which would be observable in the future. In this
paper an attempt is made to formulate a method for the
calculation, to the lowest order, of the nonlinear evolution of the 
perturbations generated from a generic multifield inflationary model. 
This is done by extending the usual perturbation theory to second
order. A future paper will address the issue from a different
perspective \cite{gp}.

\section{Perturbations and gauge invariance at second order}

Cosmological perturbation theory is a rather arcane subject. The
reason is that in a general perturbed spacetime there is no
privileged coordinate system with respect to which one can define
perturbations. So
perturbations can change when we change the coordinates. The study of
general relativistic perturbations was pioneered in \cite{lifs} and studied by many authors
since (see e.g \cite{mfb} for a comprehensive review). Let us
briefly recall a more formal presentation of what is usually meant
when one talks of perturbations in general relativity \cite{j.m.s}. One considers a five dimensional space composed of
the background spacetime ${\mathcal{M}}_0$ and, stacked above it,
perturbed spacetimes ${\mathcal{M}}_{\epsilon}$ parametrized by the parameter
$\epsilon$. We implicitly assume some sort of differentiable
structure on this 5-D space such that these perturbed spacetimes can
be considered ``close'' to ${\mathcal{M}}_0$. On these spacetimes live
tensor fields $T$. One then defines a vector field $X$, the integral
curves of which are used for identifying points on
${\mathcal{M}}_{\epsilon}$ with points on the background ${\mathcal{M}}_0$. The choice
of $X$ is completely arbitrary and is called a choice of
{\it{gauge}}. In general, any tensor $T$ can be expanded as a taylor series
\beq
T_{0}+\delta_{X}T=\phi^{*}_{X\epsilon}(T_{\epsilon})=T_0+\epsilon\pounds_XT\mid_0+\frac{1}{2}\epsilon^2\pounds_X\pounds_XT\mid_0+...\,,
\eeq
or, calling the various terms the perturbations at various orders,
\beq
\phi^{*}_{X\epsilon}(T_{\epsilon})=T_0+{\delta}T^{(1)}+{\delta}T^{(2)}+...\, ,
\eeq
where $\phi^{*}_{X\epsilon}$ is the pullback along $X$ on the
background manifold ${\mathcal{M}}_0$ of a tensor that lives on a
perturbed spacetime $\mathcal{M}_{\epsilon}$ parameter distance
$\epsilon$ away from the background. Hence the vector field $X$
alows us to define perturbations in a
meaningful way. The choice of another vector field  $Y=(1,Y^{\mu})$ defines a different gauge and
one finds that perturbations differ when defined in different
gauges:
\beq
\delta_YT-\delta_XT=\epsilon\pounds_{\xi^{(1)}}T\mid_0+\frac{1}{2}\epsilon^2\pounds_{\xi^{(1)}}^2T\mid_0+\epsilon^2\pounds_{\xi^{(1)}}\delta^{(1)}_XT+\frac{1}{2}\epsilon^2\pounds_{\xi^{(2)}}T\mid_0+...\,,
\eeq
where $\delta^{(1)}_XT\equiv\epsilon\pounds_XT\mid_0$ is the linear
perturbation of $T$ in the ``$X$ gauge'' and
$\xi^{(1)}\equiv{Y-X}$, $\xi^{(2)}\equiv{[X,Y]}$ are vector fields
which lie on ${\mathcal{M}}_0$ and are independent of
each other. Hence, by a suitable choice of ${\xi^{(1)}}$ and
${\xi^{(2)}}$, a gauge condition can be imposed order by order.

 Expansion (3) also suggests a strategy for identifying gauge invariant
quantities at second order. Observe that the first and fourth terms
on th r.h.s of (3) are essentially
the same (the transformations they define have the same functional form). This means that any linear combination $f(\delta{T}^{(1)})$ of
first order variables which is gauge invariant to first order will
also be gauge invariant w.r.t that part of the second order
transformation which corresponds to the $\frac{1}{2}\epsilon^2\pounds_{\xi^{(2)}}T\mid_0$
term in (3). The remaining terms,
$\frac{1}{2}\epsilon^2\pounds_{\xi^{(1)}}^2T\mid_0+\epsilon^2\pounds_{\xi^{(1)}}\delta^{(1)}_XT$,
are all composed of products of first order quantities. So in seeking
gauge invariant combinations at second order we must look for
appropriate quadratic terms of first order quantities that will
cancel these quadratic terms in (3). If this can be done in a unique way then
the form of a gauge invariant quantity at first order will dictate its
form at second order.

We will now give an explicit example of the construction of a second
order gauge invariant variable corresponding to the well known first
order gauge invariant quantity (first introduced in \cite{luk}, see
also \cite{mfb})
\beq
\cR=\psi+\frac{\ch}{\varphi_{0}'}\delta\varphi .
\eeq
In general every quantity will be expanded in orders like in (2). For
example 
\bea
g_{\mu\nu}&=&g^{(0)}_{\mu\nu}+{\delta}g^{(1)}_{\mu\nu}+\frac{1}{2}{\delta}g^{(2)}_{\mu\nu}+...\nonumber\\
\varphi&=&\varphi^{(0)}+{\delta}\varphi^{(1)}+\frac{1}{2}{\delta}\varphi^{(2)}+...
\eea
e.t.c. In particular, writing the general perturbed metric element as 
\beq
ds^2=a^2[(1+2\phi)d\tau^2-2B_{i}dx^id\tau-[(1-2\psi)\gamma_{ij}+2E_{ij}]dx^idx^j]
\eeq
we have
\bea
g_{00}&=&a(\tau)^2\left(1+2\phi^{(1)}+\phi^{(2)}+...\,\right)\\
g_{0i}&=&a(\tau)^2\left(B_i^{(1)}+\frac{1}{2}B_i^{(2)}+...\,\right)\\
g_{ij}&=&-a(\tau)^2\left[\left(1-2\psi^{(1)}-\psi^{(2)}+...\right)\delta_{ij}+2E_{ij}^{(1)}+E_{ij}^{(2)}+...\,\right].
\eea 
Then, from eqn. (3) one can calculate the formulae for the gauge
transformations of the relevant quantities. For an extensive account 
of second order gauge transformations, explicit formulae and some specific examples the
reader can see \cite{bruni,bruni2} and
references therein. 

In general, formulae for perturbations at second order can be
complicated and calculations rather tedious. In this paper we will make a
number of simplifying assumptions. We will ignore vectors (hence
spatially indexed quantities are given by derivatives of scalars) and, mainly,
we will drop terms containing
more than one spatial gradients. They are expected to be unimportant on
scales longer than the hubble radious. Although there is no rigorous
justification for the latter approximation it is expected to capture
the main affects on superhorizon scales \cite{long}. Within such an
approach, initial conditions at horizon crossing can be set by linear
theory. Then, the
long wavelength equations can be used to calculate the nonlinearities
induced during the superhorizon evolution. The authors of
\cite{uzan} adopted such a procedure and showed that it is possible
for significant nongaussianities to be generated in the adiabatic
mode from the long wavelength evolution in multifield inflationary
models. Their
calculation ignored metric perturbations which are included here (see
next section).      

With these approximations in mind we have the following formulae for a
gauge transformation: At first order
\bea
\tilde\phi_{(1)}&=&\phi_{(1)}+{\xi^0_{(1)}}'+\ch\xi_{(1)}^0,\\
\tilde{B}_{(1)i}&=&{B}_{(1)i}-\partial_i\xi^0_{(1)}+\xi_{i(1)}',\\
\tilde\psi_{(1)}&=&\psi_{(1)}-\ch\xi_{(1)}^0,\\
\tilde{E}_{(1)ij}&=&{E(1)}_{ij},\\
\tilde\delta\varphi_{(1)}&=&\delta\varphi_{(1)}+\varphi'\xi^0_{(1)},
\eea
and at second order 
\bea
\tilde\phi_{(2)}=\phi_{(2)}+{\xi^0_{(2)}}'+\ch\xi_{(2)}^0
&+&\xi_{(1)}^0\left[2\left(\phi_{(1)}'+2\ch\phi_{(1)}\right)+{\xi_{(1)}^0}''+5\ch{\xi_{(1)}^0}'+\left(\ch'+2\ch^2\right)\xi_{(1)}^0\right]\nonumber\\
&+&2{\xi_{(1)}^0}'\left(2\phi_{(1)}+{\xi_{(1)}^0}'\right),
\eea
\bea
\tilde{B}_{(2)i}&=&{B}_{(2)i}-4\phi_{(1)}\partial_i\xi_{(1)}^0+\xi_{(1)}^0\left[2\left({B_i^{(1)}}'+2\ch
  B_i^{(1)}\right)-\partial_i{\xi^0_{(1)}}'+{\xi_i^{(1)}}''-4\ch\left(\partial_i\xi_{(1)}^0-{\xi_i^{(1)}}'\right)\right]\nonumber\\
&+&{\xi^0_{(1)}}'\left(2B_i^{(1)}-3\partial_i\xi^0_{(1)}+{\xi_{i(1)}}'\right)+{\xi_{(1)}^j}'\left(-4\psi_{(1)}\delta_{ij}+2E_{(1)ij}\right)-\partial_i\xi^0_{(2)}+\xi_{i(2)}',
\eea
\beq
\tilde\psi_{(2)}=\psi_{(2)}-\ch\xi^0_{(2)}+\xi_{(1)}^0\left[2\left(\psi_{(1)}'+2\ch\psi_{(1)}\right)-\left(\ch'+2\ch^2\right)\xi_{(1)}^0-\ch{\xi_{(1)}^0}'\right],
\eeq
\beq
\tilde{E}_{(2)ij}={E}_{(2)ij}+2\xi^0_{(1)}\left({E}_{(1)ij}'+2\ch{E}_{(1)ij}\right)
\eeq
\beq
\tilde\delta\varphi_{(2)}=\delta\varphi_{(2)}+\varphi'\xi_{(2)}^0+\xi_{(1)}^0\left(\varphi''\xi_{(1)}^0+\varphi'{\xi_{(1)}^0}'+2\delta\varphi_{(1)}'\right).
\eeq
   Note that, as mentioned before, the part of the transformations in
(15) - (19) containing the vector field $\xi_{(2)}$ is exactly the same as the
first order case, eqn's (10) - (14). From the above we see that the variable (4) at second order transforms like
\bea
\frac{\ch}{\varphi_0'}\delta\tilde\varphi^{(2)}+\tilde\psi^{(2)}=\frac{\ch}{\varphi_0'}\delta\varphi^{(2)}+\psi^{(2)}&+&\left[\ch\frac{\varphi_0''}{\varphi_0'}-\left(\ch'+2\ch^2\right)\right](\xi_{(1)}^0)^2\nonumber\\
&+&2\left(\frac{\ch}{\varphi_0'}\delta\varphi_{(1)}'+\psi_{(1)}'+2\ch\psi_{(1)}\right)\xi_{(1)}^0
\eea 
As expected the transformation contains only products of first order
quantities. Therefore we seek to construct a gauge invariant quantity
at second order by adding a quadratic combination of first order
quantities that will transform appropriately. By inspection we see
that it must contain $\psi$, $\delta\varphi'$ and $\psi'$ and it must not contain
$\delta\varphi$. So we must have
\beq
\left(A\psi+C\psi'+D\delta\varphi'\right)\left(E\psi+G\psi'+H\delta\varphi'\right).
\eeq
Noting that 
\bea
\psi'\,\rightarrow\,\psi'-\ch'\xi^0-\ch{\xi^{0}}'\\
\delta\varphi'\,\rightarrow\,\delta\varphi'+\varphi_0''\xi^0+\varphi_0'{\xi^{0}}'
\eea
and that we must not have terms involving ${\xi^{0}}'$ we see that we have
2 options. We either set 
\bea
D=C\frac{\ch}{\varphi_0'}\\
H=G\frac{\ch}{\varphi_0'}
\eea
which eliminates the terms involving ${\xi^{0}}'$ in the transformation or
take 
\bea
D=H=\frac{4\pi}{m_p^2}\ch\varphi_0'\\
C=G=\ch^2-\ch'.
\eea
and use the background equations of motion. In both cases we are forced
to consider $A=E$, $C=G$ and $D=H$ and we end up with the same
variable \cite{mat}
\bea
\cR_{(2)}&=&\left[\psi_{(2)}+\frac{\ch}{\varphi_{0}'}\delta\varphi_{(2)}\right]\nonumber\\
&+&\frac{\left[\psi_{(1)}'+2\ch\psi_{(1)}+\frac{\ch}{\varphi_0'}\delta_{(1)}\varphi'\right]^2}{\ch'+2\ch^2-\ch\frac{\varphi_0''}{\varphi_0'}}
\eea
which is invariant under the transformations (17) and (19).

\section{Einstein equations for scalar fields at second
order}

Consider the perturbed line element
\beq
ds^2=a^2[(1+2\phi)d\tau^2-2B_{i}dx^id\tau-[(1-2\psi)\gamma_{ij}+2E_{ij}]dx^idx^j].
\eeq
Here it is understood that all quantities appearing are to be expanded
as in (5). By inserting (29) into the Einstein equations with the
relevant energy momentum tensor and keeping only linear order terms,
one arrives at the well known equations of linear perturbation
theory which can be symbolically represented as
\beq
{\mathcal{D}}\delta^{(1)}g=0\,.
\eeq
Here $\mathcal{D}$ is a set of linear differential operators and
$\delta^{(1)}g$ represents the perturbation variables. At second order
there are two types of terms. The $\delta^{(2)}g$'s
and terms quadratic in the $\delta^{(1)}g$'s. The later are supposed to
be known from the solution of the first order problem. The form of the
equations at second order will then be 
\beq
{\mathcal{D}}\delta^{(2)}g={\mathcal{J}}[(\delta{g})^2]\,.
\eeq   
with $\mathcal{D}$ the {\it{same}} operator as in (30) and
${\mathcal{J}}$ a source term quadratic in the perturbations. Since the solution to the homogeneous equation ${\mathcal{D}\delta{g}=0}$ is
known then we can consider the source terms $\mathcal{J}[(\delta{g})^2]\equiv\mathcal{J}[(\delta^{(1)}{g})^2]$ as known
functions to second order. The solution of (31) will then have the form
\beq
\delta^{(2)}g(y)={\int}{\mathcal{D}}^{-1}(y-x){\mathcal{J}}\mathrm{d}x,
\eeq
with ${\mathcal{D}}^{-1}(y-x)$ the appropriate set of Green's functions, i.e
the second order perturbations will be determined entirely by the
$\mathcal{J}$'s. If $\delta^{(1)}g$ is taken to be a gaussian random
field, then (32) shows that $\delta^{(2)}g$ is given by the square of
a gaussian random field ($\mathcal{J}$ is quadratic in first order
perturbations) and is non-gaussian.

\subsection{Gravity sector}

In the longitudinal gauge, the linear perturbation of the Einstein
tensor has the well known form (ignoring second order gradients)
\bea
\delta_LG^{0}{}_{0}&=&\frac{2}{a^2}\left[-3\ch(\ch\phi+\psi')\right],\\
\delta_LG^{0}{}_{i}&=&\frac{2}{a^2}\partial_i\left(\ch\phi+\psi'\right),\\
\delta_LG^{i}{}_{j}&=&-\frac{2}{a^2}\left((2\ch'+\ch^2)\phi+\ch\phi'+\psi''+2\ch\psi'\right)\delta^i_j\nonumber\\
&+&\frac{1}{a^2}\left[E_{ij}''+2{\ch}E_{ij}'\right],
\eea
where the subscript 'L' stands for the linear part. $E_{ij}$ is a
transverse traceless tensor (there are no vectors and we ignore the
scalar part since it is second order in spatial derivatives). For the matter
sector we will take a system of scalar fields $\varphi^A$ with the lagrangian
\beq
\mathcal{L}=\frac{1}{2}G_{AB}\partial^{\mu}\varphi^A\partial_{\mu}\varphi^B-V(\varphi)
\eeq
which corresponds to an energy momentum tensor
\beq
T_{\mu\nu}=G_{AB}\partial_\mu\varphi^A\partial_\nu\varphi^B-g_{\mu\nu}\left[\frac{1}{2}G_{AB}\partial_\lambda\varphi^A\partial^\lambda\varphi^B-V(\varphi)\right].
\eeq
Here, $V(\varphi^A)$ is an arbitrary scalar potential. The fields
$\varphi^A$ can be considered as coordinates on a field manifold with
a symmetric metric $G_{AB}$\footnote{The Lagrangian (36) is analogous
to the lagrangian for a point particle in curved space which moves
under the influence of a potential V. Here we have 4 parameters - the
4 spacetime coordinates - instead of one in the case of the point
particle.}. The background equations of motion for
the scalar fields derived
from (36) are
\beq
\frac{1}{\sqrt{-g}}\partial_\mu\left(\sqrt{-g}g^{\mu\nu}\partial_\nu\varphi^A\right)+g^{\mu\nu}\partial_\mu\varphi^B\partial_\nu\varphi^C\Gamma^A_{BC}+G^{AF}V_{,F}=0
\eeq
or 
\beq
\varphi^A{}''+2\ch\varphi^A{}'+\varphi^B{}'\varphi^C{}'\Gamma^A_{BC}+a^2G^{AF}V_{,F}=0,
\eeq
where $\Gamma^A_{BC}$ is the symmetric metric connection formed from
$G_{AB}$ 
\beq
\Gamma^A_{BC}=\frac{1}{2}G^{AD}\left(G_{BD,C}+G_{CD,B}-G_{BC,D}\right),
\eeq
and the linear perturbation of the energy momentum
tensor is
\bea
\delta_LT^0{}_0&=&\frac{1}{a^2}\left[\left(\frac{1}{2}G_{AB,C}\delta\varphi^C-{\phi}\,G_{AB}\right)\varphi^A{}'\varphi^B{}'+G_{AB}\varphi^B{}'\delta\varphi^A{}'+a^2V_{,C}\delta\varphi^C\right]\,,\\
\delta_LT^0{}_i&=&\frac{1}{a^2}G_{AB}\varphi^A{}'\partial_i\delta\varphi^B\,,\\
\delta_LT^i{}_j&=&-\frac{\delta^i{}_j}{a^{2}}\left[\left(\frac{1}{2}G_{AB,C}\delta\varphi^C-{\phi}\,G_{AB}\right)\varphi^A{}'\varphi^B{}'+G_{AB}\varphi^B{}'\delta\varphi^A{}'-a^2V_{,C}\delta\varphi^C\right].
\eea
A perturbation $\vec{\delta\varphi}$ will be a tanjent vector on the
field manifold. Given a basis $\base_A$ in field space we have 
\beq
\vec{\delta\varphi}=\delta\varphi^A\base_A.
\eeq
If the field manifold is flat there is a prefered basis, say $\{\base_1,
\base_2,...\}$ which makes the kinetic term in the lagrangian
canonical. Even in the flat case, a basis
\{$\base_A$\} can in general depend on the coordinates in field
space. We will use such a basis below. Then, $G_{AB}=\base_A\cdot\base_B\neq\delta_{AB}$. Here, $\cdot$ denotes a
product in field space.
If we expand $\phi$, $\psi$ and $\delta\varphi^A$ as in
(5) then $\delta_L$ is a linear operator acting at each order respectively. 

At second order we will also have terms from the quadratic
perturbation of the Einstein tensor and the energy momentum tensor
which we denote by $\delta_2G^{\mu}{}_{\nu}$ and
$\delta_2T^{\mu}{}_{\nu}$. Then at second order the Einstein equations
become
\bea
\delta_LG^{0}{}_{0}[\phi_{(2)},\psi_{(2)}]&=&\frac{8\pi}{m_p^2}\delta_LT^{0}{}_{0}[\phi_{(2)},\,\psi_{(2)},\,\delta\varphi^A_{(2)}]-\frac{2}{a^2}A\\
\delta_LG^{0}{}_{i}[\phi_{(2)},\psi_{(2)}]&=&\frac{8\pi}{m_p^2}\delta_LT^{0}{}_{i}[\phi_{(2)},\,\psi_{(2)},\,\delta\varphi^A_{(2)}]-\frac{2}{a^2}B_i\\
\delta_LG^{i}{}_{j}[\phi_{(2)},\psi_{(2)}]&=&\frac{8\pi}{m_p^2}\delta_LT^{i}{}_{j}[\phi_{(2)},\,\psi_{(2)},\,\delta\varphi^A_{(2)}]-\frac{2}{a^2}C^{i}{}_{j}
\eea
where
\bea
A\,&\equiv\,&\frac{a^2}{2}\left(\delta_2G^0{}_0-\frac{8\pi}{m_p^2}\delta_2T^0{}_0\right)\\
B_i\,&\equiv\,&\frac{a^2}{2}\left(\delta_2G^0{}_i-\frac{8\pi}{m_p^2}\delta_2T^0{}_i\right)\\
C^i{}_j\,&\equiv\,&\frac{a^2}{2}\left(\delta_2G^i{}_j-\frac{8\pi}{m_p^2}\delta_2T^i{}_j\right).
\eea
where, say, $\delta_2G^0{}_0$ is the quadratic perturbation of the
Einstein tensor. To this order it will contain quadratic products of
$\psi_{(1)}$ and $\delta\varphi^A_{(1)}$ which are considered known from
the solution of the linear problem. (Please note that $B_i$ here is different from the shift $B_i$ of (11) - we are in the
longitudinal gauge, and C has nothing to do with the C of the
previous section). From (34), (42) and (46) we get that 
\beq
\left(\ch\phi+\psi'\right)=\frac{4\pi}{m_p^2}G_{AB}\varphi^A{}'\delta\varphi^B-\nabla^{-2}(\partial_iB_i).
\eeq
From (35), (43) and (47) we have
\beq
\delta_LG^{(2)i}{}_i=\frac{8\pi}{m_p^2}\delta_LT^{(2)i}{}_i-\frac{2}{a^2}C
\eeq
or
\bea
(2\ch'+\ch^2)\phi+\ch\phi'&+&\psi''+2\ch\psi'=\frac{4\pi}{m_p^2}\Big[\left(\frac{1}{2}G_{AB,C}\delta\varphi^C-\phi\,G_{AB}\right)\varphi^A{}'\varphi^B{}'\nonumber\\
&+&G_{AB}\varphi^B{}'\delta\varphi^A{}'-a^2V_{,C}\delta\varphi^C\Big]-\frac{1}{3}C.
\eea
where $C=C^i{}_i$. Now use the background relation 
\beq
G_{AB}\varphi^A{}'\varphi^B{}'=\frac{m_p^2}{4\pi}(\ch^2-\ch')
\eeq
and (53) becomes
\bea
\left(\ch\phi+\psi'\right)'&+&2\ch\left(\ch\phi+\psi'\right)=\frac{4\pi}{m_p^2}\Big[\frac{1}{2}G_{AB,C}\delta\varphi^C\varphi^A{}'\varphi^B{}'\nonumber\\
&+&G_{AB}\varphi^B{}'\delta\varphi^A{}'-a^2V_{,C}\delta\varphi^C\Big]-\frac{1}{3}C.
\eea
Note that $\phi$ and $\psi$ appear only
through the combination 
\beq
\psi'+\ch\phi \equiv -\delta\ch .
\eeq
Hence, the long wavelength sector of Einstein's equations does not
contain information for the values of $\phi$ and $\psi$ separately. All
the terms containing the difference $\phi - \psi$ are dropped under
the long wavelength approximation. In linear theory it is known that $\phi
= \psi$ in the case of vanishing anisotropic stress \cite{mfb}. This
need not be the case at second order. Yet, a non vanishing value of
$\phi - \psi$ should not matter dynamically since it does not enter the equations 
explicitly. It only appears as an initial condition for $\delta\ch$ 
provided by the short wavelength part of the system. But then, from 
equations (15) and (17) we see that we can set
$\phi-\psi=0$ via a second order gauge transformation
($\xi^\mu_{(1)}=0$) by choosing 
\beq
{\xi^0_{(2)}}'+2\ch\xi^0_{(2)}=\psi_{(2)}-\phi_{(2)},  
\eeq
and keep $g_{0i}=0$ with
\beq
\xi_{i(2)}'=\partial_i\xi^0_{(2)}.
\eeq
Of course $E_{(2)ij}$ will be redefined via (18). 
Using (51) and the background equation of motion (39) we obtain in
this gauge from (55) a constraint at 
long wavelengths for the quadratic parts 
\beq
 \left(B_i\right)'+2\ch B_i={1 \over 3}\partial_i C
\eeq
Combining (45) and (47) we obtain an equation of motion for the gravitational potential
at second order 
\beq
\psi''+6\ch\psi'+2\left(\ch'+2\ch^2\right)\psi=-\frac{8\pi}{m_p^2}a^2V_{,C}\delta\varphi^C
+A-{1\over 3}C.
\eeq

\subsection{Matter sector}

Equation (60) holds for an arbitrary number of scalar fields in an arbitrary
parametrisation of the field manifold. To express
the $-\frac{8\pi}{m_p^2}a^2V_{,C}\delta\varphi^C$
term on the r.h.s it will be useful to use a basis in the scalar field
space which is `adapted' to the background field trajectory,
perturbations of which we wish to study. Such an idea was put forward in \cite{gordon} in order to
separate the entropy and adiabatic perturbations. Here we would like
to give a more geometrical flavour which can be applied to an
arbitrary number of fields and a noncanonical kinetic term. For
simplicity we study a two field example with a diagonal metric. Some
results for three fields are given in appendix B. A similar approach utilising a coordinate free language was
described in \cite{vantent} 

Consider equation (51). It is a constraint equation which relates the
evolution of the gravitational (metric) perturbations to the perturbations in
the scalar fields. In general all the fields will be linked to the
gravitational perturbations. But equation (51) holds for any choice of
coordinates on the field manifold. Therefore if we choose a set of
coordinates adapted to the background trajectory, i.e a set of
field coordinates of which only one varies along the background
trajectory and at the same time make $G_{AB}=\delta_{AB}$ on the
trajectory, then the linear term on the r.h.s of (51) will contain
only the perturbation of that field since the time derivatives of the rest will
be zero. Call $\sigma$ the coordinate that varies along the
background trajectory. It will define a basis vector
\beq
\base_{\sigma}=\frac{\partial}{\partial\sigma}=\frac{1}{\sigma'}\frac{d}{d\tau}\eeq
tangent to the background curve. Call the rest (N-1) coordinates
$s_i$. Since we want the $s_i$'s to remain constant as we move along the
trajectory we demand 
\beq
\langle\fd{}s_i,\,\base_{\sigma}\rangle=0,\,\,\,i.e\,\,\,\base_{s_i}\cdot\base_{\sigma}=0.
\eeq 
We must also impose
\beq
\langle\fd{}s_i,\,\base_{s_j}\rangle=\delta_{ij},\,\,\,i.e\,\,\,\base_{s_i}\cdot\base_{s_j}=\delta_{ij}.
\eeq
The background equations of motion in these coordinates read 
\beq
\sigma''+2\ch\sigma'+\sigma'{}^2\Gamma^{\sigma}_{\sigma\sigma}+a^2G^{\sigma{F}}V_{,F}=0
\eeq
and
\beq
\sigma'{}^2\Gamma^{s_i}_{\sigma\sigma}+a^2G^{s_iF}V_{,F}=0.
\eeq
The choices above mean that $G_{\hat{A}\hat{B}}=\delta_{\hat{A}\hat{B}}$ { \it on the
background trajectory} in the $\base_{\sigma},\,\base_{s_i}$ basis. We also observe the following: the change of
$\base_{\sigma}$ along the background trajectory - on which only
$\sigma$ varies - will be
\beq
\frac{\partial\base_{\sigma}}{\partial\sigma}\equiv{\mathcal{D}}_{\base_{\sigma}}\base_{\sigma}=\Gamma^{\hat{A}}_{\sigma\sigma}\base_{\hat{A}}.
\eeq
where ${\mathcal{D}}$ is the covariant derivative operator in
field space. Since $\base_{\sigma}$ is taken to be a vector of unit length its
variation along the trajectory will be a vector normal to it, i.e a linear
combination of the $\base_{s_i}$'s only. Therefore
$\Gamma^\sigma_{\sigma\sigma}=0$. So the background equations of
motion become
\beq
\sigma''+2\ch\sigma'+a^2\delta^{\sigma{F}}V_{,F}=0
\eeq
and
\beq
\sigma'{}^2\Gamma^{s_i}_{\sigma\sigma}+a^2\delta^{s_iF}V_{,F}=0.
\eeq
Along the trajectory we will have 
\beq
\base_{\sigma}=\Lambda^A{}_{\sigma}\base_A
\eeq
with 
\beq
\Lambda^A{}_{\sigma}=\frac{\partial\varphi^A}{\partial\sigma}=\frac{\varphi^A{}'}{\sigma'}
\eeq
We want $\base_{\sigma}$ to be a unit vector so
\beq
1=\base_{\sigma}\cdot\base_{\sigma}=G_{\sigma\sigma}=\Lambda^A{}_{\sigma}\Lambda^B{}_{\sigma}G_{AB}
\eeq
from which we get that $\sigma'=\sqrt{\varphi^A{}'\varphi^B{}'G_{AB}}$
and hence
\beq
\Lambda^A{}_{\sigma}=\frac{\varphi^A{}'}{\sqrt{\varphi^B{}'\varphi^C{}'G_{BC}}}
\eeq
The coefficients
$\Lambda^A{}_{s_i}\equiv\partial\varphi^A/\partial{s_i}$ which give the
$\base_{s_i}`s$ in terms of the original vectors
\beq
\base_{s_i}=\Lambda^A{}_{s_i}\base_A
\eeq
should be chosen such that $\base_{\sigma}\cdot\base_{s_i}=0$ and
$\base_{s_i}\cdot\base_{s_j}=\delta_{ij}$, or
\beq
G_{AB}\Lambda^A{}_{\sigma}\Lambda^B{}_{s_i}=0,\,\,G_{AB}\Lambda^A{}_{s_i}\Lambda^B{}_{s_j}=\delta_{ij}.
\eeq
For $N\geq3$ there are less equations in (74) than the number of
$\Lambda^A{}_{s_i}$'s needed to define the $N-1$ entropy vectors as we
will now show. One has $\frac{N(N-1)}{2}+N-1=\frac{N(N+1)}{2}-1$
relations from (74) and one needs $N(N-1)$ functions to determine the
$\base_{s_i}$'s in terms of the $\base_{A}$'s. So we are left with
$1+\frac{N(N-3)}{2}$ undetermined functions. In three dimensions this
is 1 which corresponds to our freedom to rotate the two isocurvature
directions in a plane normal to $\base_{\sigma}$.        

In practice it is staightforward to apply a Gram-Schmidt
orthogonalization procedure to constuct a new orthonormal basis as
follows\footnote{Many thanks to Christopher Gordon for
a stimulating discussion on this}: We already know $\base_{\sigma}$
from eqns (69) and (72). Pick any other N-1 vectors from the original basis
$\base_A=\frac{\partial}{\partial{\varphi^A}}$ which are not parallel
to $\base_{\sigma}$ (as will be the case in general). Then one has N
linearly independent vectors. By applying
the Gram-Schmidt procedure we can construct N-1
orthonormal vectors which are also normal to $\base_{\sigma}$ (see
figure 1). These
define the N-1 isocurvature (entropy) directions.\footnote{For an
  application of a Gram-Schmidt orthogonalisation to assisted
  inflation models see \cite{malik wands}} What one ends up
with is a relation between $\Lambda^A{}_{s_i}$ and
$\Lambda^A{}_\sigma$
\beq
\Lambda^A{}_{s_i}=f^A\left(\Lambda^B{}_{\sigma}\right). 
\eeq
The only thing that still needs to be specified in order for this new
coordinate system to be defined in a region around the trajectory are
derivatives w.r.t the $s_i$'s. They will be determined by demanding
that the isocurvature part of the basis,
$\base_{s_i}\equiv{\partial}/{\partial{}s_i}$, commutes with $\base_{\sigma}$
\beq
[\base_{s_i},\,\base_{\sigma}]=0.
\eeq
By acting the above commutator on the functions $\varphi^A$ we get
\beq 
\frac{\partial}{\partial{}s_i}\Lambda^A{}_{\sigma}=\frac{1}{\sigma'}\frac{\partial}{\partial\tau}\Lambda^A{}_{s_i}=\frac{1}{\sigma'}\frac{\partial
f^A}{\partial\Lambda^B_\sigma}\frac{\partial}{\partial\tau}\Lambda^B{}_\sigma,
\eeq
with $f^A$ defined in (75). The derivatives
$\partial_{s_i}\Lambda^A{}_{s_j}$ are of course defined via (77) and
(75). Equation (77) can be used to compute any derivative in the
isocurvature directions.

\begin{figure}[h]

\centerline{\epsfxsize=15cm \epsfbox{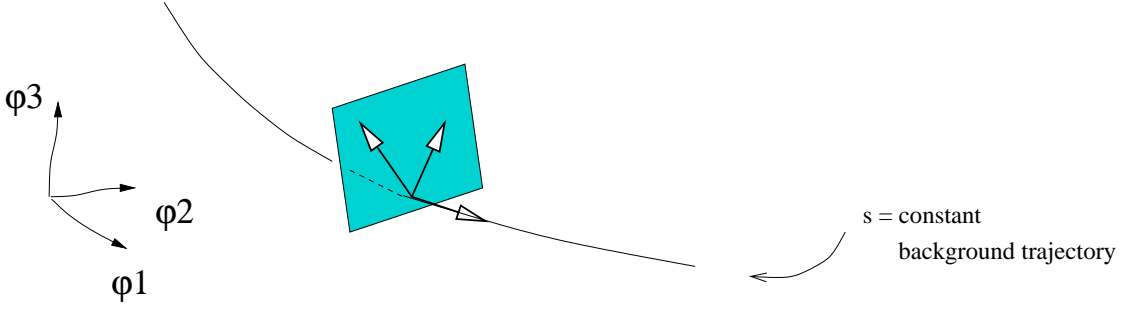}}

\caption{The new basis in N-dimensional field space. Here, $\sigma$ is a coordinate that
varies along the background trajectory. It defines a basis vector
$\base_{\sigma}$ tangent to the curve. There are $N-1$ coordinates
$s_i$ which are constant along the curve and define $N-1$ coordinate
directions normal to it. The basis vectors along these directions
$\base_{s_i}$ are taken to be orthonormal and define the $N-1$
isocurvature perturbations.}
\end{figure}

Let us now focus for simplicity on a system with two fields and a
diagonal metric 
\beq
G_{AB}=\left(\begin{array}{cc} \alpha &0 \\0 &\beta \end{array}\right).
\eeq
We take $s=0$ on the trajectory and we write (72) as 
\bea
\Lambda^1{}_{\sigma}={U}(s)\frac{\varphi^{1'}}{\sqrt{\alpha(\varphi^{1'})^2+\beta(\varphi^{2'})^2}}\\
\Lambda^2{}_{\sigma}={W}(s)\frac{\varphi^{2'}}{\sqrt{\alpha(\varphi^{1'})^2+\beta(\varphi^{2'})^2}},
\eea
where we have parametrised the dependence on $s$ by the functions
${U}(s)$ and ${W}(s)$. On the trajectory ${U}(0)={W}(0)=1$. We take
the unnormalised vector normal to the velocity to be 
\beq
\tilde\base_{s}=\base_2-(\base_{\sigma}\cdot\base_2)\base_{\sigma}=\left(1-\beta(\Lambda^1{}_{\sigma})^2\right)\base_2-\beta\Lambda^2{}_{\sigma}\Lambda^1{}_{\sigma}\base_1.
\eeq
Therefore, the isocurvature basis vector is 
\beq
\base_s=-\sqrt{\beta \over
  \alpha}\Lambda^2{}_{\sigma}\base_1+\sqrt{\alpha\over \beta}\Lambda^1{}_{\sigma}\base_2,
\eeq
from which we read off the transformation coefficients in the $s$
direction 
\beq
\Lambda^1{}_s=-\sqrt{\beta \over
  \alpha}\Lambda^2{}_{\sigma},\,\,\,\Lambda^2{}_s=\sqrt{\alpha \over
  \beta}\Lambda^1{}_{\sigma}.
\eeq
The metric in these new coordinates reads 
\beq
G_{\sigma\sigma}=G_{ss}=\alpha\left(\Lambda^1{_{\sigma}}\right)^2+\beta\left(\Lambda^2{_{\sigma}}\right)^2,
\eeq
\beq
G_{s\sigma}=0,
\eeq
which reduces to unit diagonal on the background trajectory, $s=0$, as
expected. 

In order to specify the new coordinate system completely, we need to know
the functions $U$ and $W$ appearing in (79) and (80), or,
equivalently, to be able to calculate a derivative of arbitrary order
with respect to $s$ on the background trajectory, i.e at $s=0$. This
can be achieved via (77) which in this case reads
\beq
\frac{dU}{ds}=-{W}\frac{1}{{\varphi^1}'}{\partial \over
  \partial\tau}\left(\sqrt{\beta\over \alpha}{{\varphi^2}'\over \sigma'}\right)
\eeq
and
\beq
\frac{dW}{ds}={U}\frac{1}{{\varphi^2}'}{\partial \over
  \partial\tau}\left(\sqrt{\alpha\over \beta}{{\varphi^1}'\over \sigma'}\right).
\eeq
With these, any derivative in the $\sigma, s$ coordinates can be
calculated in terms of functions of $\tau$, known along the background trajectory.

We can now derive equations for the gravitational potential and the
isocurvature perturbation at second order. The first term on the r.h.s of (60) is written in the $\sigma,\,s$
coordinates
\beq
\frac{8\pi}{m_p^2}a^2V_C\delta\varphi^C=\frac{8\pi}{m_p^2}a^2V_{,s}\delta{s}+\frac{8\pi}{m_p^2}a^2V_{,\sigma}\delta\sigma.
\eeq
From (67)
\beq
a^2V_{,\sigma}\delta\sigma=\left(\frac{\sigma''}{\sigma'}+2\ch\right)\sigma'\delta\sigma,
\eeq
and from (51)
\beq
\sigma'\delta\sigma=\frac{m_p^2}{4\pi}\left(\ch\psi+\psi'\right)+\frac{m_p^2}{4\pi}\nabla^{-2}\left(\partial_iB_i\right).\eeq
So from (60) we can get an equation for the gravitational potential to
second order
\bea 
\psi''+2\left(\ch-\frac{\sigma''}{\sigma'}\right)\psi'&+&2\left(\ch'-\ch\frac{\sigma''}{\sigma'}\right)\psi
=-\frac{8\pi}{m_p^2}\left(\sigma'\right)^2\Gamma^s_{\sigma\sigma}\delta{s}\nonumber\\
&+&A-F'-2\left({\sigma''\over \sigma'}+3\ch\right)F,
\eea
with
\beq
F\equiv\nabla^{-2}\partial_iB_i.
\eeq
and we have used that on long wavelengths we have from (59)
\beq
F'+2\ch F = {1\over 3}C.
\eeq
We also need an equation for the entropy perturbation $\delta s$. We
can get it by perturbing (38) expressed in the $\sigma,\,s$
coordinates:
\bea
\delta{}s''&+&2\ch\delta{}s'-2\psi(\sigma')^2\Gamma^{s}_{\sigma\sigma}+2\sigma'\bigg(\delta\sigma'\Gamma^{s}{}_{\sigma\sigma}+\delta{}s'\Gamma^{s}{}_{s\sigma}\bigg)\nonumber\\
&+&(\sigma')^2\bigg(\delta{}s\Gamma^{s}{}_{\sigma\sigma,s}+\delta\sigma\Gamma^{s}{}_{\sigma\sigma,\sigma}\bigg)+a^2G^{s\hat{F}}{}_{,\hat{C}}\delta\varphi^{\hat{C}}V_{,\hat{F}}+a^2V_{,s\hat{C}}\delta\varphi^{\hat{C}}=J^s.
\eea
Expressing $\delta\sigma$ and $\delta\sigma'$ in
terms of $\psi$ and $\psi'$ we finally get
\beq
\delta{}s''+2\ch\delta{}s'+\left[a^2V_{ss}-6\left(\sigma'\right)^2\Gamma^s_{\sigma\sigma}+\left(\sigma'\right)^2\partial_s\Gamma^s_{\sigma\sigma}\right]=J^s.
\eeq
The $J^s$ term appearing on the r.h.s of (95) is a second order
source and it is calculated in the appendix. Equations (91) and (95)
are generalisations to second order and a diagonal but otherwise
arbitrary field metric of the equations for the evolution of long
wavelength curvature
and isocurvature perturbations given in \cite{gordon}.   
The above procedure could of course be applied to an arbitrary number
of fields and the most general metric.

A nice result proved in \cite{gordon} is that when the background
trajectory is straight in field space then isocurvature perturbations
do not source the adiabatic one or, equivalently, the gravitational
potential perturbation. This is true only if the metric is flat,
$G_{AB}=\delta_{AB}$. If the field space has a nontrivial metric
then the equivalent statement would be that this decoupling occurs if the background
fields follow a {\it geodesic} in field space. By geodesics
we mean the curves that the background fields would follow if there
was no potential term in the lagrangian (36). We will now
prove this for the two field case but the argument can be generalised
in a straightforward manner. The geodesic equation in field space would read
\beq
\varphi^A{}''+2\ch\varphi^A{}'+\varphi^B{}'\varphi^C{}'\Gamma^A{}_{BC}=0. 
\eeq
In the $\sigma,\,s$ coordinates these equations read
\bea
\sigma''+2\ch\sigma'&=&0\nonumber\\
(\sigma')^2\Gamma^{s}{}_{\sigma\sigma}&=&0. 
\eea
Hence, such trajectories correspond to
$\Gamma^{s}{}_{\sigma\sigma}=0$ and from (91) we see that, indeed,
the isocurvature perturbations do not source the metric
perturbation \footnote{$\Gamma^{s}{}_{\sigma\sigma}$
measures the rate of change of the $\base_{\sigma}$ vector along the
trajectory in the direction of $s$. Hence it is true that a trajectory that bends generates
a coupling between isocurvature and adiabatic perturbations as
suggested in \cite{gordon}.}. In contrast, for a ``straight'' trajectory
$\Gamma^{s}{}_{\sigma\sigma}\neq0$. Indeed
\bea
\Gamma^{s}{}_{\sigma\sigma}&=&\frac{1}{2}G^{s\hat{F}}\left(2G_{\hat{F}\sigma,\sigma}-G_{\sigma\sigma,\hat{F}}\right)\nonumber\\
&=&\frac{1}{2}\left(2G_{s\sigma,\sigma}-G_{\sigma\sigma,s}\right)\nonumber\\
&=&-\frac{1}{2}G_{\sigma\sigma,s}
\eea
where we have used that, by construction,
$G_{\hat{A}\hat{B}}=\delta_{\hat{A}\hat{B}}$ along the trajectory. Now
$G_{\sigma\sigma}=\Lambda^A{}_{\sigma}\Lambda^B{}_{\sigma}G_{AB}$
so
\bea
G_{\sigma\sigma,s}&=&2\Lambda^A{}_{\sigma}\Lambda^B{}_{\sigma,s}G_{AB}+\Lambda^A{}_\sigma\Lambda^B{}_\sigma\Lambda^C{}_sG_{AB,C}\nonumber\\
&=&2G_{AB}\Lambda^A{}_{\sigma}\frac{1}{\sigma'}\frac{\partial}{\partial\tau}\Lambda^{B}{}_{s}+\Lambda^A{}_\sigma\Lambda^B{}_\sigma\Lambda^C{}_sG_{AB,C}
\eea
from eq (77). In general the 
r.h.s of (99) will be different from zero. Hence
$\Gamma^{s}{}_{\sigma\sigma}\neq0$ and isocurvature perturbations
can source the adiabatic one.

We conclude that nontrivial
kinetic terms can lead to nontrivial couplings between the adiabatic
and the isocurvature perturbations which may not be supressed on
superhorizon scales, in contrast to what happens in models with flat
field metrics \footnote{A particular example for a two field model
with a non flat metric where such a conclusion is reached was studied
in \cite{br}}. Such scalar Lagrangians appear in the effective actions of
various candidate fundamental theories. Therefore it would be
worth studying wether the interplay of isocurvature and
adiabatic perturbations in such models has any interesting
phenomenological consequences. Of course, non linear evolution will
couple the two types of perturbations anyway.

\subsection{Superhorizon Gauge Invariant Variables}
Having found solutions to the perturbation equations at second order
in a particular gauge, we can always construct gauge invariant quantities in terms of the new
field coordinates $\sigma$ and $s_i$. At first order the gauge invariant
curvature perurbation $\cR$ is can be written as \cite{gordon}
\beq
\cR_{(1)}=\left[\psi_{(1)}+\frac{\ch}{\sigma'}\delta\sigma_{(1)}\right]
\eeq
Since $s'=0$, $\delta{s_i}$ is gauge
invariant at first order. At second order, according to eqn. (19)
\beq
\tilde\delta{s}=\delta{s}+2\xi_{(1)}^0\delta{s_i}_{(1)}'
\eeq
so a corresponding gauge invariant quantity is easily seen to be 
\beq
\delta{s_i}^{(g.i)}=\delta{s_i}-2\frac{\delta\sigma_{(1)}}{\sigma'}\delta{s_i}_{(1)}'.
\eeq
Similarly we can construct a gauge invariant curvature perturbation at
second order similar to (28)
\bea
\cR_{(2)}&=&\left[\psi_{(2)}+\frac{\ch}{\sigma'}\delta\sigma_{(2)}\right]\nonumber\\
&+&\frac{\left[\psi_{(1)}'+2\ch\psi_{(1)}+\frac{\ch}{\sigma'}\delta_{(1)}\sigma'\right]^2}{\ch'+2\ch^2-\ch\frac{\sigma''}{\sigma'}}.
\eea
The variables (102) and (103) are second order gauge invariant variables
(at least on superhorizon scales) which can be used to study
isocurvature and adiabatic perurbations in the mild nonlinear regime. 

\section{Summary}
We have touched upon the issue of studying second order perturbations
in multifield inflationary models and defined gauge invariant
variables - equations (102) and (103) - on supehorizon
scales also to second order. The latter can be constructed given the
solution to lowest non linear order in a given gauge. We have presented a more geometrical method for the splitting of
isocurvature and adiabatic perturbations and applied it to a two
field  model with a diagonal but otherwise arbitrary metric. We showed
that in this case naive ``straight'' trajectories do not lead to the
decoupling of adiabatic from isocurvature perturbations. We identified
the type of
curves for which this happens. A perturbative
approach by which one can study the nonlinear evolution of
perturbations in such models to second order was suggested. The resulting equations have the same form as the first
order linear ones but with new terms appearing on the right hand side. These new terms
are quadratic in the first order perturbations and therefore they can be
considered as known ``sources'' from the solution of the first order problem. Such a formalism can be used to
calculate the amount of nongaussianity produced in such models by
treating the perturbations as gaussian stochastic fields when they become
superhorizon and then study their non linear evolution from that point. The resulting non gaussianity will be of the $\chi^2$ type
since we are considering quadratic products of gaussian fields. We
have implicitly assumed a smoothing on scales larger than the horizon
and dropped second order spatial gradients.
Although straightforward to calculate, the resulting source terms, in
principle known, are quite complicated. We will return to the issue of
calculating non linear evolution in multifield inflationary models with
a different approach in a future publication \cite{gp}.

{\bf Acknowledgements:} Many thanks to Carsten van de Bruck and Christopher
Gordon for useful discussions and comments, Marco Bruni for bringing
to my attention other works related to second order perturbation
theory, and especially Paul Shellard for discussions and generous support.

\appendix

\section{Second order sources}
In this appendix we give formulae nessecary for the calculation of the
terms A, C, F, $J^s$
defined in the text. These are the terms containing products of first
order perturbations.  

We write the metric as 
\beq
g_{\mu\nu}=g_{\mu\nu}^{B} + h_{\mu\nu}.
\eeq
where $h_{0i}=0$. Indices will be raised and lowered with the background metric. Then
the perturbation of the contravariant metric tensor will be
\beq
g^{\mu\nu}=g^{(B)\mu\nu}-h^{\mu\nu}+h^{\mu}{}_{\alpha}h^{\alpha\nu}
\eeq
and in particular
\bea
g^{00}&=&\frac{1}{a^2}-\frac{1}{a^4}h_{00}+\frac{1}{a^6}h_{00}h_{00}\\
g^{0i}&=&0\\
g^{ij}&=&-\frac{\delta_{ij}}{a^2}-\frac{1}{a^4}h_{ij}-\frac{1}{a^6}h_{ik}h_{kj}.
\eea
The perturbation of the metric determinant is
\beq
\sqrt{-g}=a^4\left(1+\frac{1}{2}h+\frac{1}{8}h^2-\frac{1}{4}h^{\alpha}{}_{\beta}h^{\beta}{}_{\alpha}\right).
\eeq
The perturbation of the Riemann tensor to first and second order can be found to
be (\cite{mtw} p:965)
\beq
R^{(1)}_{\mu\nu}=\frac{1}{2}\left(-h_{|\mu\nu}-h_{\mu\nu|\alpha}{}^\alpha+h_{\alpha\mu|\nu}{}^\alpha+h_{\alpha\nu|\mu}{}^\alpha\right)
\eeq
and
\bea
R^{(2)}_{\mu\nu}&=&\frac{1}{2}\bigg[\frac{1}{2}h_{\alpha\beta{|}\mu}h^{\alpha\beta}{}_{|\nu}+h^{\alpha\beta}\left(h_{\alpha\beta|\mu\nu}+h_{\mu\nu|\alpha\beta}-h_{\alpha\mu|\nu\beta}-h_{\alpha\nu|\mu\beta}\right)\nonumber\\
&+&h_{\nu}{}^{\alpha|\beta}\left(h_{\alpha\mu|\beta}-h_{\beta\mu|\alpha}\right)-\left(h^{\alpha\beta}{}_{|\beta}-\frac{1}{2}h^{|\alpha}\right)\left(h_{\alpha\mu|\nu}+h_{\alpha\nu|\mu}-h_{\mu\nu|\alpha}\right)\bigg].
\eea
where a vertical bar denotes a covariant derivative w.r.t. the
background. Then
\beq
R^{(1)\mu}{}_{\nu}=g^{(B)\mu\alpha}R^{(1)}{}_{\alpha\nu}-h^{\mu\alpha}R^{(B)}{}_{\alpha\nu}
\eeq
and
\beq
R^{(2)\mu}{}_{\nu}=g^{(B)\mu\alpha}R^{(2)}_{\alpha\nu}-h^{\mu\alpha}R_{(1)\alpha\nu}+h^{\mu}{}_{\lambda}h^{\lambda\alpha}R^{(B)}_{\alpha\nu}.
\eeq  
Using the following perturbed metric tensor (longitudinal gauge)
\beq
h_{00}=2a^2\psi
\eeq
\beq
h_{ij}=2a^2\psi\delta_{ij}-2a^2E_{ij}
\eeq
\beq
h=-4\psi
\eeq
we get, after a rather tedious calculation
\beq
R^{(2)0}{}_{0}=\frac{2}{a^2}\bigg[-6\ch\psi'\psi-6\left(\ch'+2\ch^2\right)\psi^2-E_{kl}\left(\ch{E_{kl}'}+2\ch^2E_{kl}\right)+\frac{1}{2}E'_{kl}E'_{kl}\bigg],
\eeq
\beq
R^{(2)0}{}_i=\frac{2}{a^2}\bigg[\partial_i\psi\left(\psi'-4\ch\psi\right)+2E_{kl}\left(\partial_iE_{kl}'-\partial_lE_{ki}'\right)
+E_{kl}'\partial_iE_{kl}+2E_{il}\partial_l\psi'-2\partial_l\psi{}E_{il}'\bigg],
\eeq
\bea
R^{(2)i}{}_j&=&-\frac{4}{a^2}\bigg[\psi\left(\ch\psi'-(\ch'+2\ch^2)\psi\right)\delta_{ij}+\frac{1}{2}(\psi')^2\delta_{ij}\nonumber\\
&+&\frac{1}{2}{E}_{ij}\left(\psi''+2\ch\psi'+2(4\ch'+9\ch^2)\psi\right)\nonumber\\
&-&\frac{1}{2}\ch{}E'_{ik}E_{kj}-\frac{1}{2}\ch{}E'_{kl}E_{kl}\delta_{ij}-\frac{1}{4}(\ch'+2\ch^2)E_{ik}E_{kj}\bigg].
\eea
The Einstein tensor will be
\beq
G^{(2)0}{}_0=\frac{1}{2}\left(R^{(2)0}{}_0-R^{(2)l}{}_l\right)
\eeq
\beq
G^{(2)0}{}_i=R^{(2)0}{}_i
\eeq
\beq
G^{(2)i}{}_j=R^{(2)i}{}_j-\frac{1}{2}\delta^i{}_j\left(R^{(2)0}{}_0-R^{(2)i}{}_i\right),
\eeq
so we find
\bea
G^{(2)0}{}_0&=&\frac{1}{a^2}\bigg[-6\psi\left((\ch'+2\ch^2)\psi\right)+3(\psi')^2\nonumber\\
&-&E_{kl}\left(5\ch{}E_{kl}'+(\frac{9}{2}\ch^2-\ch')E_{kl}\right)
+\frac{1}{2}E_{kl}'E_{kl}'\bigg],
\eea
\bea
G^{(2)0}{}_i&=&\frac{2}{a^2}\bigg[\partial_i\psi\left(\psi'-4\ch\psi\right)+2E_{kl}\left(\partial_iE_{kl}'-\partial_lE_{ki}'\right)\nonumber\\
&+&E_{kl}'\partial_iE_{kl}+2E_{il}\partial_l\psi'-2\partial_l\psi{}E_{il}'\bigg],
\eea
\bea
G^{(2)i}{}_{j}&=&\frac{3}{a^2}\bigg[2\psi\left(4\ch\psi'+2(\ch'+2\ch^2)\psi\right)\delta_{ij}+(\psi')^2\delta_{ij}\nonumber\\
&+&2\ch{}E_{ki}'E_{kj}-\ch{}E_{kl}'E_{kl}\delta_{ij}+(\ch'+2\ch^2)E_{ki}E_{kj}\nonumber\\
&-&\frac{1}{2}(\ch^2-2\ch')E_{kl}E_{kl}\delta_{ij}
-\frac{1}{2}E'_{kl}E'_{kl}\delta_{ij}\nonumber\\
&-&2E_{ij}\left(\psi''+2\ch\psi'+2(4\ch'+9\ch^2)\psi\right)\bigg].
\eea
For the energy momentum tensor 
\beq
T_{\mu\nu}=G_{AB}\partial_\mu\varphi^A\partial_\nu\varphi^B-g_{\mu\nu}\left[\frac{1}{2}G_{AB}\partial_\lambda\varphi^A\partial^\lambda\varphi^B-V(\varphi)\right].
\eeq
the second order perturbation for two fields is given by 
\bea
T^{(2)0}{}_{0}&=&\frac{1}{a^2}\bigg[\left(\delta\sigma'\right)^2-2\sigma'\left(\Gamma^s_{\sigma\sigma}\right)'\delta
  s\delta\sigma-\left(\sigma'\right)^2\partial_s\Gamma^s_{\sigma\sigma}\delta s^2\nonumber\\
&+&2\psi^2\left(\sigma'\right)^2+2\left(\sigma'\right)^2\Gamma^s_{\sigma\sigma}\psi\delta s-2\Gamma^s_{\sigma\sigma}\sigma'\delta s\delta\sigma'-2\sigma'\psi\delta\sigma'\bigg]\nonumber\\
&+&V_{ss}\delta s^2+2V_{s\sigma}\delta s\delta\sigma+V_{\sigma\sigma}\delta\sigma^2
\eea
\beq
T^{(2)0}{}_i=\frac{1}{a^2}\left[\delta{}s'\partial_i\delta{}s+\left(\delta\sigma'-2\sigma'\Gamma^s_{\sigma\sigma}\delta{}s-2\psi\sigma'\right)\partial_i\delta\sigma\right],
\eeq
\bea
T^{(2)i}{}_{j}&=&-\delta^i_j\frac{1}{a^2}\bigg[\left(\delta\sigma'\right)^2-2\sigma'\left(\Gamma^s_{\sigma\sigma}\right)'\delta
  s\delta\sigma-\left(\sigma'\right)^2\partial_s\Gamma^s_{\sigma\sigma}\delta s^2\nonumber\\
&+&2\psi^2\left(\sigma'\right)^2+2\left(\sigma'\right)^2\Gamma^s_{\sigma\sigma}\psi\delta
  s-2\Gamma^s_{\sigma\sigma}\sigma'\delta
  s\delta\sigma'-2\sigma'\psi\delta\sigma'\bigg]\nonumber\\
&+&\delta^i_j\left[V_{ss}\delta s^2+2V_{s\sigma}\delta s\delta\sigma+V_{\sigma\sigma}\delta\sigma^2\right]
\eea
where we have obviously dropped second order spatial derivatives. We also need to compute the quadratic part of the wave equation for
the system of scalar fields
\beq
\frac{1}{\sqrt{-g}}\partial_\mu\left(\sqrt{-g}g^{\mu\nu}\partial_\nu\varphi^A\right)+g^{\mu\nu}\partial_\mu\varphi^B\partial_\nu\varphi^C\Gamma^A_{BC}+G^{AF}V_{,F}=0.
\eeq
Up to second order spatial gradients we get 
\bea
J^s&=&-4{d\over d\tau}\left(a^2\psi\delta
s'\right)+4a^2\left(\sigma'\right)^2\Gamma^s_{\sigma\sigma}\left(\psi^2+\Gamma^s_{\sigma\sigma}\psi\delta
s\right)\nonumber\\
&+&a^2\sigma'\left(\Gamma^s_{\sigma\sigma}\right)'\left[-3\psi\delta\sigma+2{\delta\sigma\delta\sigma'\over
    \sigma'}+2{\delta\sigma\delta s'\over
    \sigma'}+4\Gamma^s_{\sigma\sigma}\delta
  s\delta\sigma\right]\nonumber\\
&+&a^2\partial_s\Gamma^s_{\sigma\sigma}\left[-2\delta
  s^2-3\left(\sigma'\right)^2\psi\delta s+2{\sigma'\delta s\delta\sigma'}+2{\sigma'\delta s\delta s'}\right]\nonumber\\
&+&a^2\left(\sigma'\right)^2\delta\sigma\left[{1\over\sigma'}{d \over
    d\tau}\left({1\over
    \sigma'}\left(\Gamma^s_{\sigma\sigma}\right)'\right)+{1\over\sigma'}{d \over d\tau}\left(\partial_s\Gamma^s_{\sigma\sigma}\right)\right]\nonumber\\
&-&2a^4\psi\left(V_{ss}\delta s
+V_{s\sigma}\delta\sigma\right)+a^4\left(V_{sss}\delta{}s^2+2V_{ss\sigma}\delta{}s\delta\sigma+V_{s\sigma\sigma}\delta\sigma^2\right)\bigg]\nonumber\\
&+&a^2\left(\sigma'\right)^2\Gamma^s_{\sigma\sigma}E_{ij}E_{ij}.
\eea
With these formulae we can compute all the second order terms defined
in the text.

\section{Isocurvature-adiabatic split with three fields}

In this appendix we present the isocurvature-adiabatic split
for three fields $\varphi^1,\,\varphi^2,\,\varphi^3$ with an
arbitrary metric $G_{AB}$. This should illustrate the general case. Now we will have two isocurvature
directions. Take the set $\{\base_{\sigma},\,\base_2,\,\base_3\}$ as
the starting point for constructing the orthonormal vectors. Form
$\base_{s_2}$ which lies in the plane spanned by $\base_{\sigma}$ and
$\base_{2}$. First form the unnormalised
\beq
\tilde\base_{s_2}=\base_2-(\base_2\cdot\base_{\sigma})\base_{\sigma}
\eeq
One finds 
\beq
\tilde\base_{s_2}=\tilde\Lambda^A{}_{s_2}\base_A
\eeq
with
\bea
\tilde\Lambda^2{}_{s_2}&=&1-\Lambda_{2\sigma}\Lambda^2{}_{\sigma}=\Lambda_{1\sigma}\Lambda^1{}_{\sigma}+\Lambda_{3\sigma}\Lambda^3{}_{\sigma}\nonumber\\
\tilde\Lambda^1{}_{s_2}&=&-\Lambda_{2\sigma}\Lambda^1{}_{\sigma}\nonumber\\
\tilde\Lambda^3{}_{s_2}&=&-\Lambda_{2\sigma}\Lambda^3{}_{\sigma},
\eea
where
\beq
\Lambda_{A\sigma}\equiv{}G_{AB}\Lambda^B{}_{\sigma}.
\eeq
Then
\beq
\base_{s_2}=\frac{\tilde\Lambda^A{}_{s_2}}{\sqrt{\tilde\Lambda^B{}_{s_2}\tilde\Lambda_{Bs_2}}}\base_A.
\eeq
So, we have 
\beq
\Lambda^A{}_{s_2}=\frac{\tilde\Lambda^A{}_{s_2}}{\sqrt{\tilde\Lambda^B{}_{s_2}\tilde\Lambda_{Bs_2}}}.
\eeq
Repeating the procedure we form
\beq
\tilde\base_{s_3}=\base_3-(\base_3\cdot\base_{\sigma})\base_{\sigma}-(\base_3\cdot\base_{s_2})\base_{s_2}
\eeq
to get
\bea
\tilde\Lambda^1{}_{s_3}&=&-\Lambda_{3\sigma}\Lambda^1{}_{\sigma}-\Lambda_{3s_2}\Lambda^1{}_{s_2}\nonumber\\
\tilde\Lambda^2{}_{s_3}&=&-\Lambda_{3\sigma}\Lambda^2{}_{\sigma}-\Lambda_{3s_2}\Lambda^2{}_{s_2}\nonumber\\
\tilde\Lambda^3{}_{s_3}&=&1-\Lambda_{3\sigma}\Lambda^3{}_{\sigma}-\Lambda_{3s_2}\Lambda^3{}_{s_2}.
\eea
So we have 
\beq
\Lambda^A{}_{s_3}=\frac{\tilde\Lambda^A{}_{s_3}}{\sqrt{\tilde\Lambda^B{}_{s_3}\tilde\Lambda_{Bs_3}}}.
\eeq
As before, we demand for the $\base_{s_i}$`s
\beq
\left[{\base_{s_i}},\,\base_{\sigma}\right]=0 
\eeq
from which we get by acting on $\varphi^A$
\bea
\frac{1}{\sigma'}\frac{\partial}{\partial\tau}\Lambda^A{}_{s_2}=\frac{\partial}{\partial{}s_2}\Lambda^A{}_{\sigma}\\
\frac{1}{\sigma'}\frac{\partial}{\partial\tau}\Lambda^A{}_{s_3}=\frac{\partial}{\partial{}s_3}\Lambda^A{}_{\sigma}.
\eea
Knowing the $\Lambda$ coefficients we can calculate the metric $G_{\hat{A}\hat{B}}$ in the
new basis, where the index $\hat{A}$ runs over
$\{\sigma,\,s_1,\,s_2\}$. Using equations (142) and (143) we can then calculate any partial derivative of
$\Lambda^A{}_{\sigma}$ in terms of time derivatives along the
background trajectory. Hence we can calculate the connection and
perturbations of arbitrary order in the new
$\{\base_{\sigma},\,\base_{s_1},\,\base_{s_2}\}$ basis.

\end{document}